# Versatile Lithium Niobate Platform for Photoacoustic/Thermoelastic Gas Sensing and Photodetection

Haoyang Lin, Wenguo Zhu, Yongchun Zhong, Huihui Lu, Jianhui Yu, and Huadan Zheng*

*Department of Optoelectronic Engineering, Jinan University, Guangzhou, 510632, China*

*Email correspondance : zhenghuadan@jnu.edu.cn*

We present a lithium niobate multi-functional platform (LN-MFP) that integrates photoacoustic and thermoelastic spectroscopy with photodetection on a single chip. Utilizing LN's piezoelectric and thermoelastic properties and an optimized design, it achieves high sensitivity across visible to long-wave infrared wavelengths. We demonstrate photoacoustic/thermoelastic gas sensing and photodetection using LN-MFP. The compact, integrated system with custom packaging and direct wire bonding reduces complexity and size, enabling portable, scalable sensing for environmental and diagnostic applications.

Lithium niobate (LN) has emerged as a key material for on-chip devices, thanks to its outstanding electro-optic, piezoelectric, and nonlinear optical properties [1]. These unique characteristics make it a versatile platform for diverse integrated photonic applications [2]. Recent breakthroughs in LN photonics have propelled the development in high-performance components, including electro-optic modulators [3-5], frequency shifters [6], frequency combs [7-9], pulse generators [10], tunable lasers [11], and photonic integrated circuits [12]. Such advancements are critical enablers for optical communications, signal processing, and high-sensitivity sensing technologies [13, 14]. Moreover, LN's distinctive piezoelectric and thermoelastic properties unlock new possibilities for high-performance acoustic and acousto-optic devices, expanding its range of applications [15-17]. The advancement of LN-based on-chip devices is pivotal to the future of integrated photonics and electromagnetic spectrum technologies [18]. LN's exceptional versatility positions it as a cornerstone for the next generation of compact, high-performance, and energy-efficient photonic systems.

Advancements in spectroscopic methodologies are expanding the frontiers photonics, offering unprecedented insights into molecular structures and dynamics. These breakthroughs are transforming fields such as chemistry, physics, and environmental science [19-23]. Among the spectroscopic methods, photoacoustic spectroscopy (PAS) stands out for its non-destructive nature and versatility across diverse type of samples over an extended range of wavelengths [24-28]. Light-induced thermoelastic spectroscopy (LITES) is an optical gas sensing technique that detects target gases by converting the sensor's thermoelastic deformation, induced by modulated light absorption, into electrical signals. This enables contact-based detection with high sensitivity and selectivity, particularly when using materials with excellent piezoelectric properties [29-31].

Despite their individual strengths, integrating spectroscopic detectors remains a major challenge for advancing micro-, nano- and on-chip spectroscopic devices. Conventional systems often rely on bulky optical components and complex configurations, limiting their suitability for miniaturized and portable sensing applications [32]. Key obstacles include large instrumentation, high power consumption, and the precise alignment of optical components, which hinder field deployment and real-time, on-site analysis. Furthermore, lack of seamless integration across multiple spectroscopic techniques restricts the development of versatile, compact platforms capable of delivering comprehensive analytical insights.

To overcome these challenges, this work introduces lithium niobate as a multi-functional integrated platform (LN-MFP) for spectroscopic sensing. By harnessing LN's strong piezoelectric and thermoelastic properties, we develop an integrated sensor capable of performing diverse spectroscopic measurements within a single device. The LN-MFP seamlessly combines photoacoustic, thermoelastic, and photodetection functionalities, addressing the limitations of conventional spectroscopic systems. This integration not only minimizes physical footprint and system complexity but also enhances sensitivity and selectivity through the synergistic interplay of multiple detection mechanisms. Incorporating resonance mechanisms can significantly enhance sensor performance, improving both sensitivity and selectivity across diverse applications [33]. In this work, we leverage lithium niobate's piezoelectric properties and thermoelastic effects to detect and amplify signals via mechanical resonance, further optimized through a fork-shaped structure. This work has the potential to revolutionize on-chip spectroscopic devices by providing a compact, high-performance, and versatile solution that that overcomes the limitations of conventional bulky systems. The LN-MFP not only allows to advance the fundamental understanding of LN-based integrated platforms but also lays the groundwork for future innovations in the spectroscopic technologies. By addressing key integration challenges and enhancing spectroscopic detection capabilities, this research marks a pivotal step toward the next generation of multifunctional, on-chip spectroscopic sensors, driving progress in integrated photonics and advanced sensing technologies.

Fig. 1(a) presents the design of the LN-MFP. Simulations performed with COMSOL

Multiphysics software provided a detailed analysis of the electrical response of lithium niobate under varying stress and thermal conditions. The simulation results in Fig. 1(b), depict the charge distribution across the LN-MFP during vibration. Positive charge accumulations appear in the red regions, while negative charges are shown in the blue regions. Notably, negative charges are concentrated near the gap between the tines, whereas positive charges accumulate near the device's outer boundaries, forming a distinct bipolar distribution. This pattern suggests that mechanical deformation or thermal effects enhance the piezoelectric response, particularly in the marginal regions. Moreover, the symmetric charge distribution across both tines underscores their synchronized motion during vibration.

Fig. 1(c) outlines the eight-step fabrication process of the LN-MFP. It begins with slicing a y-cut 128° lithium niobate crystal block into thin wafers, each measuring 500 μm in thickness and 4 inches in diameter. The wafer edges are then refined using circular grinding to eliminate remove surfaces and ensure dimensional uniformity. Next, dual rotating pads polish the wafers to achieve a smooth, uniform finish. Finally, a spray-type cleaning system removes any contaminants and residues, preparing the wafer for subsequent processing. After cleaning, both sides of the wafers are coated with a 0.2 μm gold film. The wafers are then precisely diced into individual components. Fig. 1(c) illustrates the structural design of the LN-MFP, which is fabricated using mechanical polishing and laser processing. The tines of the LN-MFP are ~ 11.5 mm in length and 1.7 mm in width, with a gap of 1 mm between them. The base has a width of W~ 7.6 mm, and features a circular fillet with a radius of r~ 0.8 mm at the junction between the base and the tines.

The design of the LN-MFP was guided by two primary factors: resonance frequency and the efficient coupling with laser sources. To minimize 1/f noise during spectroscopic detection, controlling the resonance frequency is crucial. In photoacoustic and photo-thermoelastic detection, the modulation frequency must be moderate, as certain molecular non-radiative relaxation processes can persist for up to 100 microseconds. Therefore, optimizing the resonance frequency is essential to maximizing device sensitivity while minimizing noise. Additionally, light sources like terahertz (THz) quantum cascade lasers (QCLs) [34] or Fabry-Perot diode lasers [35] often exhibit suboptimal beam quality. To accommodate these sources, the gap between the tines is intentionally widened. Microelectrodes were fabricated on both sides of the device using magnetron sputtering and masking techniques.

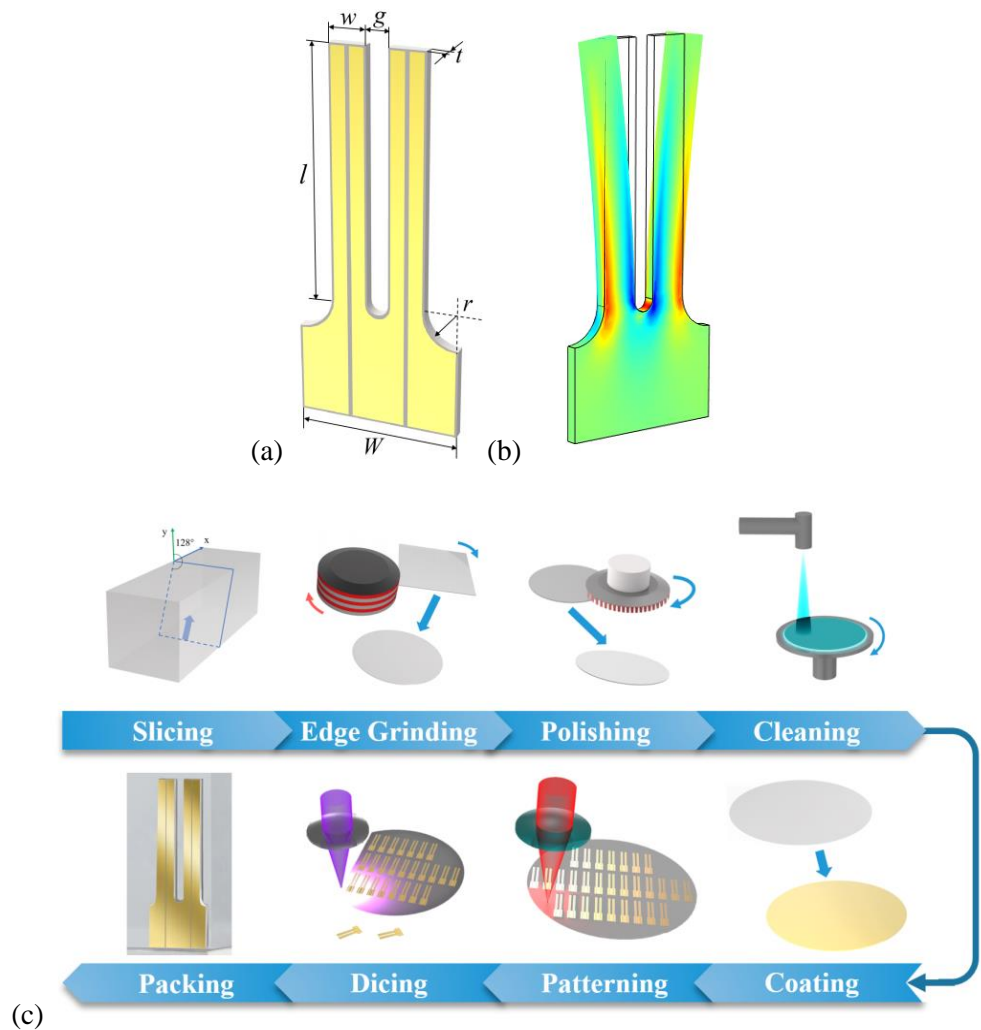

Fig. 1 Lithium Niobate Multi-Functional Platform. (a) Design of the LN-MFP. (b) Simulated surface charge distribution of LN-MFP during vibration. The red regions correspond to positive charge accumulations, blue regions to negative ones. (c) The manufacturing process of LN-MFP.

To assess the photoacoustic detection capabilities of the LN-MFP, we employed light sources of different types covering a broad spectral range, from visible to long-wave infra-red wavelengths, specifically at 450 nm, 1.3 µm, 1.5 µm, 2 µm, 3.3 µm, and 9.77 µm, shown in the Fig. 2(a). This wide spectral operating range enables the detection of various gas species by targeting their characteristic absorption lines. All experiments were carried out at room temperature and pressure.

Nitrogen dioxide ($NO_2$) detection is crucial for environmental monitoring and public health protection. To measure $NO_2$ concentration, we used a visible (VIS) blue LED with a center wavelength of 450 nm as the light source. The target gas, supplied at a certified concentration of 5 ppm in $N_2$, was analyzed, and the results are shown in Fig. 2(b). Since the LED operated at a fixed wavelength, intensity modulation was applied to excite $NO_2$ molecules and generate acoustic waves. The LED emitted an output power of ~ 1.2 W, operating at a modulation frequency of 10,485 Hz with a 50% duty cycle. The LN-MFP achieved a signal-to-noise ratio (SNR) of ~132, corresponding to a minimum detection limit (MDL) of ~38 ppb.

The concentration of water vapor ($H_2O$) in air was measured using a 1.3 µm near-infrared (NIR) distributed feedback (DFB) semiconductor laser, with the results presented in Fig. 2(c). The laser wavelength was scanned from 7185.2 $cm^{-1}$ to 7186.1 $cm^{-1}$, covering the water absorption line at 7185.6 $cm^{-1}$. At a water vapor concentration of 18,000 ppm, the peak voltage of the second harmonic signal ($2f$) was recorded at 0.0618 V. The $1\sigma$ standard deviation at a non-absorption wavelength was determined to be $3.65 \times 10^{-6}$ V, resulting in a $H_2O$ MDL of ~1 ppm.

The detection of acetylene ($C_2H_2$) is essential for ensuring safety in industrial processes and monitoring human activities. Fig. 2(d) illustrates the detection of 50 ppm acetylene in $N_2$ using a 1.5 µm NIR semiconductor laser. The laser wavelength was scanned from 6506.99 $cm^{-1}$ to 6507.63 $cm^{-1}$, covering the $C_2H_2$ absorption line at 6507.4 $cm^{-1}$. The peak voltage of the $2f$ signal was recorded at 0.0477 V, while the modulation depth was to 0.44 $cm^{-1}$. As a result, the $1\sigma$ standard deviation at a non-absorption wavelength was $5.99 \times 10^{-6}$ V, yielding a MDL of ~63 ppb for $C_2H_2$.

Carbon dioxide ($CO_2$), a greenhouse gas closely linked to climate change, was detected using a 2 μm NIR DFB laser diode. A certified $CO_2$ gas mixture with a concentration of 1000 ppm in $N_2$ was analyzed, as shown in Fig. 2(e). The laser wavelength was scanned from 4988.53 cm$^{-1}$ to 4988.77 cm$^{-1}$, covering the $CO_2$ absorption line at 4988.65 cm$^{-1}$. The 2$f$ peak signal was recorded at $1.34\times10^{-3}$ V, with a 1σ standard deviation of $1.14\times10^{-5}$ V measured away from the absorption line. These results indicate that the LN-MFP achieved a MDL of ~9 ppm for $CO_2$.

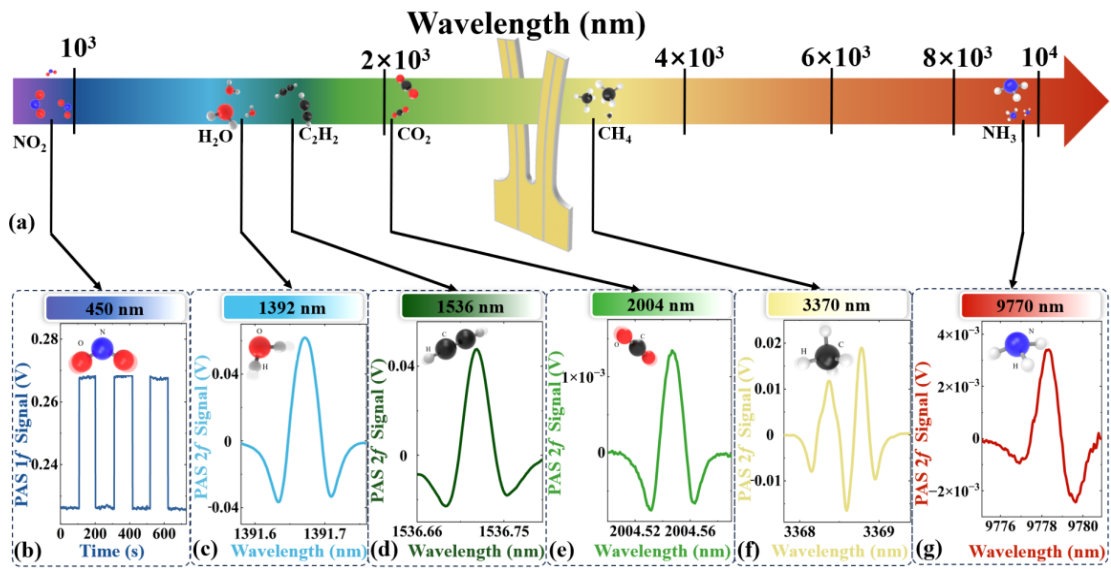

Fig. 2 LN-MFP for photoacoustic detection. (a) Overview of the detection wavelengths ranging from visible light to longwave infrared. Representative QEPAS spectra: (b) For 5 ppm $NO_2$ in $N_2$ using a 450 nm VIS LED. (c) For 18,000 ppm $H_2O$ in $N_2$ using a 1.3 μm NIR LD. (d) For 50 ppm $C_2H_2$ in N2 using a 1.5 μm NIR LD. (e) For 1000 ppm $CO_2$ in $N_2$ using a 2 μm NIR LD. (f) For 100 ppm $CH_4$ in $N_2$ using a 3.3 μm MWIR ICL. (g) For 1 ppm $NH_3$ in $N_2$ using a 9.77 μm LWIR QCL. VIS LED: visible light emitting diode. NIR LD: Near Infra-Red Laser Diode. MWIR ICL: Mid-Wave Infra-Red Interband Cascade Laser. LWIR QCL: Long-Wave Infra-Red Quantum Cascade Laser..

Methane ($CH_4$), a major greenhouse gas and widely used energy source, was analyzed using a 3.3 μm mid-wave infrared (MWIR) interband cascade laser (ICL) in combination with the LN-MFP. The measurement results are shown in Fig. 2(f). The ICL wavelength was scanned from

2967.86 cm$^{-1}$ to 2969.33 cm$^{-1}$, covering multiple methane absorption lines. Methane exhibits four absorption lines at 2968.86 cm$^{-1}$, 2968.73 cm$^{-1}$, 2968.47 cm$^{-1}$, and 2968.4 cm$^{-1}$. A certified methane gas sample with a concentration of 100 ppm in N$_2$ was used for the measurement. At the strongest absorption peak, the signal voltage reached 0.019 V, with a 1σ standard deviation of 9.4×10$^{-6}$ V measured away from the methane absorption lines. This corresponds to a MDL of ~49 ppb.

Ammonia (NH$_3$), a critical compound in both industrial and agricultural applications, was analyzed using a 9.77 μm long-wave infrared (LWIR) QCL. The results are shown in Fig. 2(g). The laser wavelength was tuned from 1022.4 cm$^{-1}$ to 1023 cm$^{-1}$, covering the ammonia absorption line at 1022.76 cm$^{-1}$. A certified ammonia gas sample with a concentration of 1 ppm in N$_2$ was used for the measurement. The system achieved a SNR of ~145, corresponding to a MDL of ~7 ppb for NH$_3$.

Table 1 Summary of LN-MFP photoacoustic detection performance.

| Light sources | Wavelength | Molecules | Power (mW) | Modulation amplitude | Sensitivity | NNEA (cm$^{-1}$·W·Hz$^{-1/2}$) |
|---|---|---|---|---|---|---|
| VIS LED | 450 nm | NO$_2$ | 1200 | / | 38 ppb | / |
| NIR LD | 1.39 μm | H$_2$O | 19 | 0.86 cm$^{-1}$ | 1 ppm | 5.82×10$^{-9}$ |
|  | 1.53 μm | C$_2$H$_2$ | 3700 | 0.44 cm$^{-1}$ | 63 ppb | 2.7×10$^{-8}$ |
|  | 2 μm | CO$_2$ | 10.6 | 0.27 cm$^{-1}$ | 9 ppm | 2.53×10$^{-8}$ |
| MWIR ICL | 3.37 μm | CH$_4$ | 6.5 | 0.29 cm$^{-1}$ | 49 ppb | 8.36×10$^{-9}$ |
| LWIR QCL | 9.77 μm | NH$_3$ | 82.3 | 0.14 cm$^{-1}$ | 7 ppb | 6.58×10$^{-10}$ |

Table 1 provides summary of the sensitivities achieved together with the calculated normalized noise equivalent absorption coefficient (NNEA) values and the optimal modulation amplitude. These results demonstrate the LN-MFP's capability to detect multiple gas species across a broad spectral range with high sensitivity, underscoring its versatility and potential for practical applications not limited to environmental monitoring and industrial process control.

Beyond its photoacoustic detection capabilities, the LN-MFP also exhibits strong potential for LITES. In this mechanism, target molecules absorb modulated laser radiation, producing localized heating and subsequent thermal expansion of a piezoelectric sensing element, in our case the LN-

MFP. This resulting mechanical deformation, enhanced by the LN-MFP's inherent mechanical resonance, excites a piezoelectric response, generating a measurable electrical signal. Notably, this resonant amplification effect significantly boosts the sensitivity of the system.

To evaluate the LITES performance of the LN-MFP system, six different light sources spanning from VIS to LWIR were employed to analyze different gases molecules and the results are shown in Fig. 3 (a). The laser beams propagated in free space, with the LN-MFP positioned at an optical path length of 2.5 cm. All experiments were conducted under laboratory conditions at ambient temperature and pressure, with $N_2$ as the carrier gas.

A 450 nm visible LED was used to detect $NO_2$. The 1f signal was obtained by alternately introducing pure $N_2$ and 5% $NO_2$, as shown in Fig. 3(b). The modulation depth was set to ~100%, with a duty cycle of 50%. The LED output power was set as 330 mW. By analyzing the signals for pure $N_2$ and 5% $NO_2$, the LN-MFP system achieved a SNR of 3,692, corresponding to a MDL of ~13.5 ppm.

For the evaluation of the thermoelastic detection capability of the LN-MFP in the near- and mid-IR regions, the noise level was determined as the 1σ standard deviation at non-absorbing wavelengths.

At 1.3 μm a near-infrared distributed feedback (DFB) laser was employed to target the $H_2O$ absorption line at 7185.6 cm$^{-1}$, with the concentration fixed at 2% via a humidity controller. Under an optimized modulation depth of 0.4 cm$^{-1}$, the LN-MFP system generated a 2$f$ LITES signal peak of 1.62 mV, as shown in Fig. 3(c), yielding a minimum detection limit (MDL) of ~59.3 ppm for $H_2O$. Similarly, a 1.5 μm near-infrared telecommunication laser was employed to detect $C_2H_2$ by targeting its absorption line at 6534.6 cm$^{-1}$. With the laser power set to 10.5 mW and a modulation depth of 0.68 cm$^{-1}$, the LN-MFP produced a 2$f$ signal peak of 4.4 mV, while the noise resulted 9.16 μV, leading to an MDL of 10.4 ppm for $C_2H_2$ as depicted in Fig. 3(d).

A 2 μm laser diode was employed to $CO_2$ at a 20% concentration. The laser was precisely tuned to target the CO$_2$ absorption line at 4988.65 cm$^{-1}$, and with a modulation amplitude of 0.47 cm$^{-1}$, the LN-MFP system generated a 2$f$ signal peak of 3.1 mV, shown in the Figure 3(e). The baseline noise was determined to be 1.1×10$^{-5}$ V, corresponding to a minimum detection limit (MDL) of ~ 680 ppm for CO$_2$. At 3.3 μm, a mid-wave infrared (MWIR) interband cascade laser (ICL) at was used to measure CH$_4$ at a concentration of 2%. Fig. 3(f) depicts the 2f signal resulting from four CH$_4$ absorption lines located at 2968.86 cm$^{-1}$, 2968.73 cm$^{-1}$, 2968.47 cm$^{-1}$, and 2968.4 cm$^{-1}$ combined. The modulation depth was optimized to 0.27 cm$^{-1}$, yielding a signal peak of 0.101 V. The 1σ noise level was determined to be 112 μV, corresponding to a MDL of ~ 22 ppm for CH$_4$.

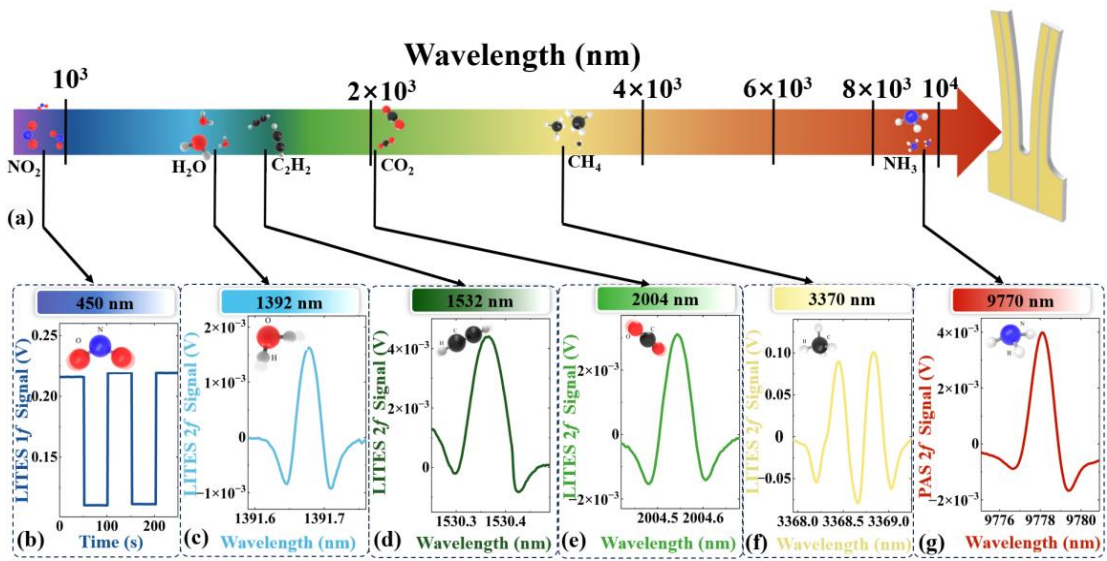

Fig. 3 LN for thermoelastic detection. (a) Overview of the detection wavelengths ranging from visible light to longwave infrared. LITES spectra measured for: (b) 5% NO$_2$ in N$_2$ using a 450 nm VIS LED; (c) 2% H$_2$O in N$_2$ using a 1.3 μm NIR LD; (d) 5,000 ppm C$_2$H$_2$ in N$_2$ using a 1.5 μm NIR LD; (e) 20% CO$_2$ in N$_2$ using a 2 μm NIR LD; (f) 2% CH$_4$ in N$_2$ using a 3.3 μm MWIR ICL; (g) 5% NH$_3$ in N$_2$ using a 9.77 μm LWIR QCL.

Finally to evaluate the thermoelastic detection capability of the LN-MFP at long-wave infrared region, a 9.7 μm quantum cascade laser (QCL) was employed for ammonia (NH$_3$) detection. The QCL was tuned from 1022.4 cm$^{-1}$ to 1023 cm$^{-1}$, covering the NH3 absorption line at 1022.76 cm$^{-1}$. With an optimized modulation depth of ~0.15 cm$^{-1}$ for a 5% NH$_3$ sample, the LN-MFP achieved a SNR of 2014, yielding a MDL of ~25 ppm. The results were shown in the Fig. 3(g).

Table 2 provides an overview of the LN-MFP's thermoelastic detection performance for various gas species across the range 450 nm - 9.77 µm. The table details the light sources type, the target molecules, the laser power, the modulation amplitude, the sensitivity, and the normalized noise-equivalent absorption (NNEA) achieved. Note that we implemented a very short optical path of only 2.5 cm and the sensitivity and NNEA could be simply improved by employing multi-pass cells or resonant cavities.

The integration of thermoelastic detection into the LN-MFP demonstrates its versatility as a multifunctional spectroscopic platform. The capability to perform both photoacoustic and thermoelastic measurements within a single device highlights its potential for comprehensive spectroscopic analysis. Such dual functionality proves especially valuable for applications requiring both high sensitivity and spectral selectivity across a broad spectral range, further positioning the LN-MFP as an advanced tool for next-generation spectroscopic sensing technologies.

Table 2 Summary of LN-MFP thermoelastic detection performance.

| Light sources | Wavelength | Molecules | Power (mW) | Modulation amplitude | Sensitivity | NNEA ($cm^{-1} \cdot W \cdot Hz^{-1/2}$) |
|---|---|---|---|---|---|---|
| VIS LED | 450 nm | $NO_2$ | 330 | / | 13.5 ppm | / |
| NIR LD | 1.39 µm | $H_2O$ | 19 | 0.4 $cm^{-1}$ | 59.3 ppm | $3.45 \times 10^{-7}$ |
| | 1.53 µm | $C_2H_2$ | 9.8 | 0.68 $cm^{-1}$ | 10.4 ppm | $2.33 \times 10^{-7}$ |
| | 2 µm | $CO_2$ | 8.5 | 0.47 $cm^{-1}$ | 680 ppm | $2 \times 10^{-6}$ |
| MWIR ICL | 3.37 µm | $CH_4$ | 6.5 | 0.27 $cm^{-1}$ | 22 ppm | $3.75 \times 10^{-6}$ |
| LWIR QCL | 9.77 µm | $NH_3$ | 82.3 | 0.15 $cm^{-1}$ | 25 ppm | $2.35 \times 10^{-6}$ |

Beyond its capabilities in photoacoustic and thermoelastic detection, the LN-MFP demonstrates remarkable potential for photodetection across a broad spectral range, from visible light at 450 nm to long-wave infrared radiation at nearly 10 µm. To assess its photodetection performance, square-wave modulated lasers at various wavelengths were shined onto the LN-MFP most sensitive area.

The modulated light intensity was gradually reduced to evaluate the device's response, enabling the determination of its minimum detectable power and overall detectivity. Fig. 4 illustrates the response curves of the LN-MFP measured for light wavelength ranging from 450 nm to 9.77 μm.

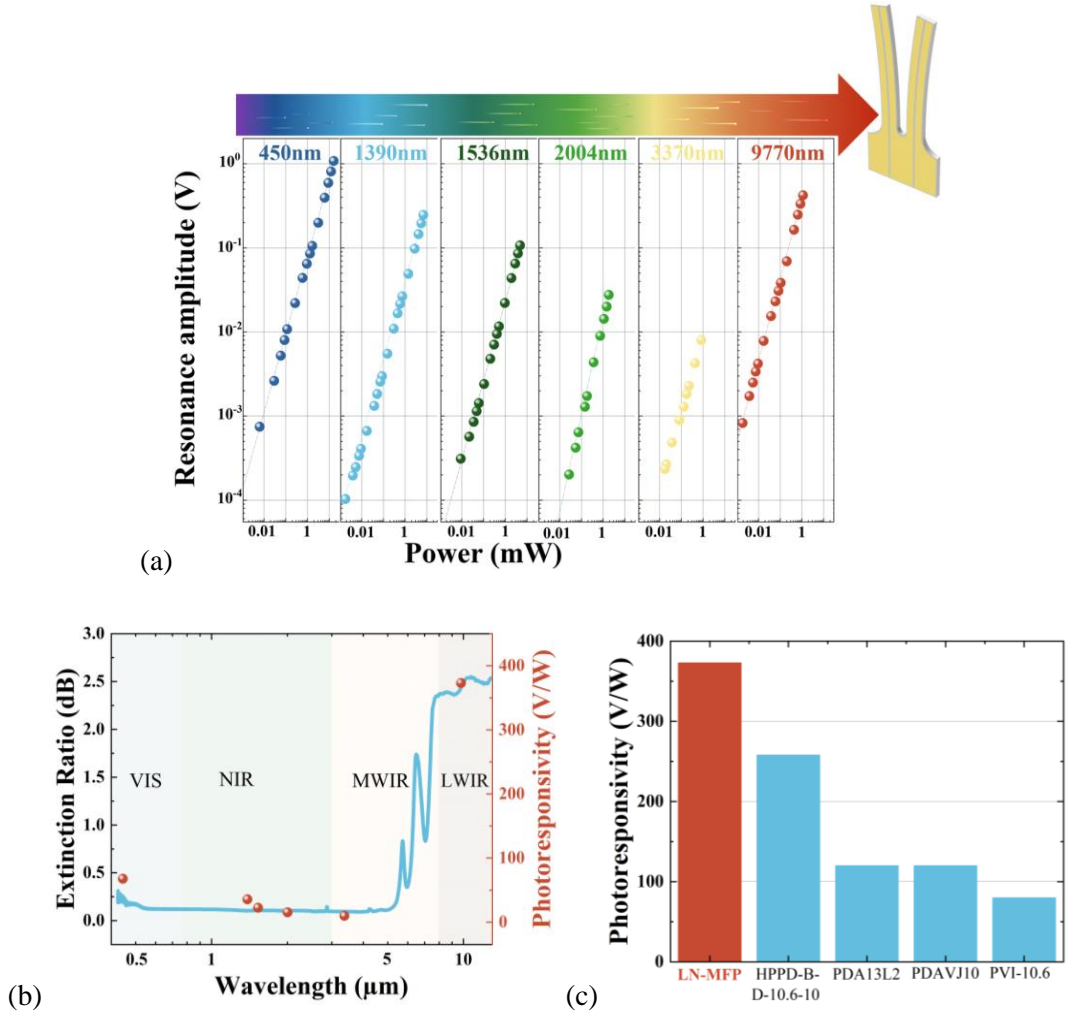

Fig. 4 LN-MFP for photodetection. (a) Resonance amplitude of LN-MFP to 450 nm VIS LED, 1.39 μm NIR LD, 1.53 μm NIR LD, 2 μm NIR LD, 3.37 μm MWIR ICL and 9.77 μm LWIR QCL. (b) The LN-MFP photoresponsivity to wavelengths from VIS to LWIR, compared to FTIR absorption spectrum of a LN wafer. (c) Photoresponsivity of LN-MFP compared with four commercial detectors at 9.7 μm, including Healthy Photo infrared detector model HPPD-B-D-10.6-10 (HgCdTe), Thorlabs pyroelectric detector model PDA13L2 (LiTaO3), photoelectric detector model PDAVJ10 (HgCdTe), Vigo infrared detector model PVI-10.6 (HgCdTe). The results of LN-MFP and HPPD-B-D-10.6-10 were obtained by experiments. The results of PDA13L2, PDAVJ10 and PVI-10.6 were obtained by the datasheet.

For the 450 nm LED as the modulation depth decreased from 0.904 $cm^{-1}$ to $3.57\times10^{-4}$ $cm^{-1}$,

the vibration amplitude of the LN-MFP correspondingly dropped from 1.08 V to $7.47\times10^{-4}$ V. The sensitivity of the LN-MFP to 450 nm light was determined to be 260 nW, with a corresponding detectivity of 67.7 V/W. The diode laser operating at 1.3 µm was detuned from the $H_2O$ absorption peak to avoid any power absorption occurring along the optical pathlength between laser and photodetector. For the 1.3 µm wavelength, as the modulation depth decreased from 1.15 $cm^{-1}$ to $2.98\times10^{-4}$ $cm^{-1}$, the vibration amplitude correspondingly dropped from $2.47\times10^{-1}$ V to $1.03\times10^{-4}$ V. The sensitivity was determined to be 445 nW, with a detectivity of 35.5 V/W. Similarly, for the 1.53 µm wavelength, decreasing the modulation depth from 0.59 $cm^{-1}$ to $1.12\times10^{-3}$ $cm^{-1}$ led to a reduction in vibration amplitude from $1.07\times10^{-1}$ V to $3.1\times10^{-4}$ V. The sensitivity of the LN-MFP to the 1.53 µm laser was then determined to be 695 nW, with a corresponding detectivity of 22.5 V/W. For the 2 µm NIR semiconductor laser, when the modulation depth was lowered from 0.33 $cm^{-1}$ to $4.95\times10^{-3}$ $cm^{-1}$, the vibration amplitude simultaneously decreased from $2.76\times10^{-2}$ V to $2.02\times10^{-4}$ V. Under these conditions, the sensitivity was measured at 693 nW, and the detectivity reached 15.3 V/W. At 3.37 µm the laser modulation depth was reduced from 0.34 $cm^{-1}$ to $7.22\times10^{-3}$ $cm^{-1}$ and the vibration amplitude decreased from $8.03\times10^{-3}$ V to $2.35\times10^{-4}$ V. The sensitivity was calculated to be 1 µW, with a corresponding detectivity of 9.8 V/W. Finally, at 9.77 µm as the modulation depth decreased from $2.47\times10^{-2}$ $cm^{-1}$ to $8.25\times10^{-5}$ $cm^{-1}$, the vibration amplitude dropped from 0.42 V to $1.73\times10^{-3}$ V. The sensitivity to the 9.77 µm laser was determined to be 25 nW, with a detectivity of $3.73\times10^{2}$ V/W.

Fig. 4(b) provides a comprehensive analysis of the LN-MFP's performance across the VIS, NIR, MWIR and LWIR spectral regions, covering wavelengths from 450 nm to 9.77 µm. The results show that the photoresponsivity initially decreases as the wavelength transitions from the VIS to the LWIR region, followed by a significant increase in the LWIR region. This trend indicates that the LN-MFP achieves its highest sensitivity in the LWIR range. This behavior is attributed to the intrinsic optical absorption properties of LN in the investigated range, as illustrated in Fig. 4(b), which also includes the absorption spectrum of LN, which was obtained by measuring the absorption of a 500 µm thick LN wafer using Fourier transform infrared spectroscopy. The enhanced absorption in the MIR and LWIR regions further underscores the LN-MFP's potential for high-sensitivity applications, such as the fundamental absorption band of molecules. Fig. 4(c) compares the photoresponsivity of the LN-MFP to that of four commercial mid-infrared detectors in response to a 9.7 µm laser. Notably, the LN-MFP exhibits superior photoresponsivity at 9.7 µm, outperforming these commercial mid-infrared photodetectors.

Table 3 LN-MFP for photodetection from VIS to LWIR.

|  | Light sources | MDL | Wavelength | Photoresponsivity (V/W) |
|---|---|---|---|---|
| LN-MFP | VIS LED | 260 nW | 450 nm | 67.7 |
|  | NIR laser diode | 445 nW | 1.39 µm | 35.5 |
|  |  | 695 nW | 1.53 µm | 22.5 |
|  |  | 693 nW | 2 µm | 15.3 |
|  | MIR interband cascade laser | 1 µW | 3.37 µm | 9.8 |
|  | LWIR quantum cascade laser | 25 nW | 9.77 µm | 373 |
| HPPD-B-D-10.6-10 | LWIR quantum cascade laser | / | 9.77 µm | 258 |

Table 3 summarizes the LN-MFP's photodetection performance across various wavelengths. All measurements were conducted at room temperature without active cooling. Notably, the minimum detection level and response rate could be significantly enhanced by incorporating cooling strategies, as lower temperatures generally reduce thermal noise and improve detector performance. This suggests that further optimization through thermal management could unlock even greater sensitivity for advanced spectroscopic applications.

Based on the abovementioned multifunctional capabilities, we introduced an designed on-chip packaging approach and evaluated its performance. We realized a custom-designed 3.5 cm×4.5 cm printed circuit board (PCB) with integrated transimpedance amplifier function to pre-process the charge output of LN-MFP and a QCL chip. The PCB features a stepped electrode pad to interface with the bottom-side electrode of the LN-MFP, while the top electrode is wire-bonded for signal extraction.

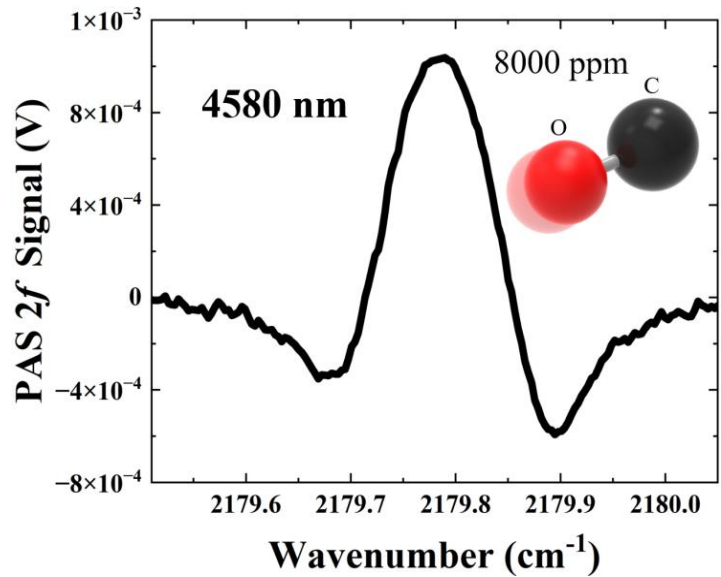

Fig. 5 Spectral signal measured from 8000 ppm CO in $N_2$.

As proof of concept experiment, a QCL chip with the central wavelength of 4.6 µm was used. The output wavelength of this QCL chip was tuned by injection current to cover the CO absorption line at 2179.77 $cm^{-1}$. The LN-MFP is coupled to the QCL chip with a distance of ~100 µm. We have initially optimized the modulation amplitude to 0.36 $cm^{-1}$. We employed a trumpet-shaped gas interface that directly covers the chip. The gas is fed to the sensor through this interface, flowing onto the chip at a rate of 100 sccm—consistent with standard experimental conditions—to help minimize noise. Fig. 5(b) presents the 2$f$ signal of 8000 ppm CO in $N_2$ measured using the packaged LN-MFP-QCL chip, demonstrating the platform's strong potential for fully integrated on-chip spectroscopic applications.

The development of the lithium niobate multi-functional platform represents a significant breakthrough in integrated spectroscopic sensing technologies. By harnessing the multifunctional properties of LN and incorporating a resonant fork-shaped design, we have successfully integrated photoacoustic spectroscopy, light induced thermoelastic spectroscopy, and photodetection into a single compact device. Experimental results reveal high sensitivity and selectivity across a broad spectral range, achieving detection limits suitable for real-world applications in environmental monitoring and industrial process control. This platform effectively addresses the integration challenges associated with conventional spectroscopic systems, providing a scalable and versatile solution that enhances performance while significantly reducing size and complexity, and system requirements.

In future work, we aim to achieve full photonic integration by combining the LN-MFP with an on-chip light source to develop a completely integrated optical sensor chip. Specifically, our goal is to merge a mid-infrared quantum cascade laser (QCL) chip with the LN-MFP using advanced chip fabrication techniques, thereby consolidating both the light source and the detector onto a single platform. This seamless integration will enable simultaneous mass production of the entire system on one chip, further miniaturizing the sensor and enhancing its functionality. Such a breakthrough is expected to open new avenues for applications in point-of-care diagnostics, real-time environmental monitoring, and in-situ chemical analysis. Beyond extending the capabilities of lithium niobate-based integrated platforms, our work represents a significant stride toward multifunctional on-chip sensors, ushering in a new era in integrated photonics and spectroscopic technologies.

Conflict of interest

The authors declare that there are no conflicts of interest.

Acknowledgment
This work is supported by the National Key Research and Development Program of China (2021YFB2800801), National Natural Science Foundation of China (62375111, 62005105, 12174156, 12174155, 62105125, 62175137), the Ministry of Education of China (8091B03012309), Natural Science Foundation of Guangdong Province (2023B1515020027), the Science and Technology Projects of Guangzhou (202102020445), Special Project in Key Fields of the Higher Education Institutions of Guangdong Province (2020ZDZX3022)